\documentclass[12pt,nofootinbib]{article}

\usepackage{graphicx}
\usepackage{amssymb}
\usepackage{amsmath}
\usepackage{color}
\usepackage[colorlinks=true,linkcolor=blue,citecolor=blue]{hyperref}

\begin{document}
\newcommand{\rem}[1]{{\bf [#1]}} \newcommand{\gsim}{ \mathop{}_
{\textstyle \sim}^{\textstyle >} } \newcommand{\lsim}{ \mathop{}_
{\textstyle \sim}^{\textstyle <} } \newcommand{\vev}[1]{ \left\langle
{#1} \right\rangle } \newcommand{\bear}{\begin{array}} \newcommand
{\eear}{\end{array}} \newcommand{\bea}{\begin{eqnarray}}
\newcommand{\eea}{\end{eqnarray}} \newcommand{\beq}{\begin{equation}}
\newcommand{\eeq}{\end{equation}} \newcommand{\bef}{\begin{figure}}
\newcommand {\eef}{\end{figure}} \newcommand{\bec}{\begin{center}}
\newcommand {\eec}{\end{center}} \newcommand{\non}{\nonumber}
\newcommand {\eqn}[1]{\beq {#1}\eeq} \newcommand{\la}{\left\langle}
\newcommand{\ra}{\right\rangle} \newcommand{\ds}{\displaystyle}
\newcommand{\red}{\textcolor{red}} 
\def\SEC#1{Sec.~\ref{#1}} \def\FIG#1{Fig.~\ref{#1}}
\def\EQ#1{Eq.~(\ref{#1})} \def\EQS#1{Eqs.~(\ref{#1})} \def\lrf#1#2{
\left(\frac{#1}{#2}\right)} \def\lrfp#1#2#3{ \left(\frac{#1}{#2}
\right)^{#3}} \def\GEV#1{10^{#1}{\rm\,GeV}}
\def\MEV#1{10^{#1}{\rm\,MeV}} \def\KEV#1{10^{#1}{\rm\,keV}}
\def\REF#1{(\ref{#1})} \def\lrf#1#2{ \left(\frac{#1}{#2}\right)}
\def\lrfp#1#2#3{ \left(\frac{#1}{#2} \right)^{#3}} \def\OG#1{ {\cal
O}(#1){\rm\,GeV}}

\begin{titlepage}

\begin{flushright}
IPMU15-0136 
\end{flushright}

\begin{center}

\vskip 1.2cm

{\usefont{T1}{ppl}{m}{n}
{\Large Can thermal inflation be consistent with baryogenesis \\[0.5em]
in gauge-mediated SUSY breaking models?} 
}

\vskip 1.2cm

{\usefont{T1}{bch}{m}{n}
Taku Hayakawa$^a$, 
Masahiro Kawasaki$^{a,b}$
and
Masaki Yamada$^{a,b}$}

\vskip 0.4cm

{\it$^a$Institute for Cosmic Ray Research, The University of Tokyo,
5-1-5 Kashiwanoha, Kashiwa, Chiba 277-8582, Japan}\\
{\it$^b$Kavli IPMU (WPI), UTIAS, The University of Tokyo, 5-1-5 Kashiwanoha, 
Kashiwa, 277-8583, Japan}\\

\date{\today}

\vskip 1.2cm

\begin{abstract}
Thermal inflation is an attractive idea to dilute cosmic density of unwanted particles
such as moduli fields
which cause cosmological difficulties.
However,
it also dilutes preexisting baryon asymmetry and
some viable baryogenesis is necessary for a cosmologically consistent scenario.
We investigate whether the Affleck-Dine mechanism can produce 
baryon asymmetry enough to survive after the dilution
in gauge-mediated SUSY breaking models.
Flat directions except for $LH_u$ flat direction cannot provide
such huge baryon number because of Q-ball formation.
We show that although the $LH_u$ flat direction is special in terms of 
having $\mu$-term which prevents Q-ball formation, 
it cannot explain the observed baryon asymmetry either.
\end{abstract}

\end{center}
\end{titlepage}

\baselineskip 6mm

\section{Introduction}
\label{sec:intro}

Supersymmetry (SUSY)~\cite{Nilles:1983ge} is one of the most attractive candidates
for extensions of the Standard Model.
It can not only naturally solve the hierarchy problem but also 
achieve the unification of the gauge couplings,
which implies the existence of Grand Unified Theories (GUTs).
Because no supersymmetric partners have been discovered so far,
however,
SUSY must be spontaneously broken,
which generally causes a SUSY flavor problem.
Gauge-mediated SUSY breaking models~\cite{Dine:1993yw,Dine:1994vc,Dine:1995ag}
give a solution to the flavor problem
since the mediation by the gauge interactions
leads to the degeneracy of sfermion masses.
We focus on the gauge-mediated SUSY breaking models in this paper.

In terms of the cosmology of SUSY models,
some long-lived particles may cause serious problems.
Local SUSY predicts the existence of gravitino, which is the superpartner of the graviton.
Although initial abundance of gravitinos is diluted during the primordial inflation, 
they are produced again through scattering process 
and decay from sparticles during reheating.
In gauge-mediation models,
the gravitino mass $m_{3/2}$ is much lighter than the electroweak scale
($m_{3/2}\ll {\cal O}(100){\rm\,GeV}$).
It interacts with other particles only through gravitational interaction 
and hence has a long lifetime.
In particular,
if the lifetime of the gravitino is larger than the present age of the universe,
which is actually the case for $m_{3/2}\lesssim{\cal O}(100){\rm\,MeV}$,
its abundance may give too much contribution to 
the present cosmic density of the universe~\cite{Moroi:1993mb,de Gouvea:1997tn}.
In order to avoid this problem called ``the gravitino problem",
the reheating temperature should be low enough 
since the produced gravitino abundance is proportional to the reheating temperature.

In the framework of superstring theories,
the situation gets worse.
There exists a lot of singlet scalar fields corresponding to flat directions in superstring theories.
We call them ``moduli fields".
The moduli fields generally acquire mass comparable to the gravitino mass through some non-perturbative effects of the SUSY breaking~\cite{de Carlos:1993jw}.
They also interact with other particles only through gravitation 
(i.e., Planck suppressed operators)
and hence have long lifetimes.
Let us focus on one of the moduli fields and consider its dynamics in the early universe.
During and after the inflation,
vacuum energy which causes the expansion of the universe largely breaks SUSY and lifts the moduli potential.
Then,
the moduli field acquires mass of the order of the Hubble parameter
and sits down at the minimum of the potential 
which is in general different from the one at present.
When the Hubble parameter becomes of the order of the moduli mass,
the moduli field starts to oscillate around the true minimum.
Its initial amplitude is expected to be of the order of the Planck scale
because the gravitational scale is the only scale appearing in supergravity actions.
A large number of moduli particles are then produced as coherent oscillations.
The moduli field oscillates until present and their energy density leads to overclosure
in the case of $m_{\eta}\lesssim{\cal O}(100){\rm\,MeV}$,
where $m_{\eta}$ is moduli mass.
If the moduli field has mass of $m_{\eta}\sim0.1{\rm\,MeV}\mathchar`-1{\rm\,GeV}$,
it generally decays into X-rays during the present epoch and
it may give too much contribution to the X-ray background spectrum, 
which more severely constrains the abundance of the moduli~\cite{Kawasaki:1997ah,Hashiba:1997rp}.
This cosmological difficulty is sometimes called 
``the cosmological moduli problem"~\cite{de Carlos:1993jw,Coughlan:1983ci,Banks:1993en}.
An important point is that the moduli problem cannot be solved even if 
the reheating temperature is as low as ${\cal O}(10){\rm\,MeV}$, 
which is restricted from below to realize the Big Bang nucleosynthesis (BBN).
The moduli problem is therefore more serious than the gravitino problem.

In order to solve the moduli problem,
late-time entropy production is needed to dilute the moduli abundance.
A probable candidate for a huge entropy dilution mechanism is ``thermal inflation"~\cite{Yamamoto:1985rd,Lyth:1995ka}
which is mini-inflation caused by a scalar field called ``a flaton".
Introducing the flaton is also motivated in light of a solution to the $\mu$ problem~\cite{Kim:1983dt,Stewart:1996ai}:
why is the $\mu$-term electroweak scale and so small compared to the GUT or Planck scale?
After the thermal inflation, 
flaton decay can produce a large amount of entropy enough to solve the moduli problem.
Indeed,
it has been studied that some specific models succeed
in solving the moduli problem~\cite{Asaka:1997rv,Asaka:1999xd}.

However, 
one is faced with another problem incidental to the entropy dilution.
The entropy production dilutes
not only the moduli abundance but also baryon number
which may be produced beforehand by some mechanisms.
In particular,
the thermal leptogenesis~\cite{Fukugita:1986hr,Buchmuller:2004nz}
cannot produce such huge baryon number that it can explain the observed baryon asymmetry after the dilution.
Hence,
we need a more effective baryogenesis
or a mechanism to generate baryon number after the entropy production.
One of the most probable candidates for the former mechanisms
is the Affleck-Dine mechanism~\cite{Affleck:1984fy,Dine:1995kz},
which can produce huge baryon asymmetry 
via dynamics of a flat direction in the visible sector.
If a flat direction carrying baryon charge, called ``the Affleck-Dine field",
acquires a nonzero vacuum expectation value (VEV) in the early universe,
it starts to oscillate after the primordial inflation
and rotates in the complex plane,
which represents the generation of baryon number.

The baryon number production by the Affleck-Dine mechanism
is closely related to Q-ball~\cite{Coleman:1985ki} formation.
Q-ball is a non-topological soliton that may be 
formed during the oscillation of the Affleck-Dine field~\cite{Kusenko:1997si,Enqvist:1998en,Kasuya:1999wu}.
It is well known that the Affleck-Dine field fragments into Q-balls
if its potential becomes flatter than a quadratic term for a larger VEV.
Since Q-balls absorb (almost) all produced baryon charge~\cite{Kasuya:1999wu,Kasuya:2000sc,Kasuya:2001hg},
we need to calculate baryon number released from these Q-balls
to estimate the baryon asymmetry in the quark sector.

In this paper, we study if the Affleck-Dine baryogenesis is compatible with the thermal inflation
in gauge-mediated SUSY breaking models.
The potentials of the Affleck-Dine fields except for the $LH_u$ flat direction
become flat at large VEVs in gauge-mediation models~\cite{de Gouvea:1997tn},
which implies that Q-balls are formed after the Affleck-Dine baryogenesis.
Unfortunately,
it has been revealed
that Q-ball formation rendered the Affleck-Dine mechanism very unlikely to work 
in the presence of the moduli problem~\cite{Kasuya:2001tp}.
In this paper,
we refine the estimated baryon number in~\cite{Kasuya:2001tp}
using a new lower bound of the SUSY breaking scale,
and show that this scenario is more unlikely to work.
On the other hand,
the $LH_u$ flat direction is special in terms of
having the $\mu$-term which prevents it from forming into Q-balls.
Although one can easily find that in the simplest case the resulting baryon asymmetry is not
much enough to explain the observed abundance,
we provide another scenario that provides larger amount of baryon asymmetry.
Motivated to solve the $\mu$ problem~\cite{Kim:1983dt,Stewart:1996ai},
we assume that the flaton couples with Higgs supermultiplets
and its VEV generates the $\mu$-term at low energy.
In this case,
the $LH_u$ flat direction starts to oscillate by a potential without the $\mu$-term,
which implies that it forms into Q-balls.
Then,
after the thermal inflation ends,
the $\mu$-term is generated and the $LH_u$ flat direction obtains the $\mu$-term,
which results in Q-ball deformation.
We study if the released lepton number
can explain the present baryon asymmetry
in this scenario.

The organization of this paper is as follows.
In Sec.~\ref{sec:moduli},
we briefly explain the moduli problem
and the thermal inflation.
First,
we consider a model of the thermal inflation introducing a flaton supermultiplet with a discrete symmetry.
In this model,
however,
flaton decay into long-lived $R$-axions could lead to a cosmological difficulty
in gauge-mediated SUSY breaking models.
In order to prohibit the decay channel into $R$-axions,
we introduce a symmetry breaking term
as proposed in~\cite{Asaka:1997rv,Asaka:1999xd}.
In Sec.~\ref{sec:AD_Q-ball},
we explain the Affleck-Dine mechanism as a candidate for an efficient production mechanism of baryon asymmetry.
We also review Q-ball and its properties.
In Sec.~\ref{sec:moduli_Affleck},
we show that the Affleck-Dine baryogenesis
is incompatible with the moduli problem
when Q-balls form after the Affleck-Dine baryogenesis.
Then,
we study if the $LH_u$ flat direction can explain the baryon asymmetry
in Sec.~\ref{sec:LH_u}.
The final section is devoted to summary and discussions.

\section{Moduli problem and Thermal inflation}
\label{sec:moduli}

In this section,
we review the moduli problem and the thermal inflation in gauge-mediated SUSY breaking models.

\subsection{Moduli problem}
\label{subsec:moduli}

In gauge-mediation models,
the moduli mass is much lighter than the electroweak scale, $m_{\eta}\ll{\cal O}(100){\rm\,GeV}$,
since it acquires mass of the order of the gravitino mass
through Planck-suppressed interactions.
If the moduli mass is lighter than ${\cal O}(100){\rm\,MeV}$,
its lifetime is larger than the present age of the universe.
Therefore, the moduli energy density must be smaller than the dark matter density,
$\rho_{\rm DM}\simeq 0.26\rho_{\rm cr}$,
where $\rho_{\rm DM}$ and $\rho_{\rm cr}$ denote the present dark matter density
and the critical density,
respectively.
The ratio of the critical density to the present entropy density is given by
\begin{equation}
	\frac{\rho_{\rm cr}}{s_0}\simeq 3.6\times10^{-9}h^2{\rm\,GeV},
	\label{eq:critical_density}
\end{equation}
where $s_0$ is the present entropy density and
$h$ denotes the present Hubble parameter in units of $100{\rm\,km/sec\,Mpc}^{-1}$.
Furthermore, 
the energy density of moduli with mass $m_{\eta}\gtrsim200{\rm\,keV}$ is more severely
constrained from the observation of X-ray background spectra
if the moduli field decays into photons via a coupling with the kinetic terms of the gauge fields~\cite{Kawasaki:1997ah,Hashiba:1997rp}.
Hereafter,
we focus on the case of $m_{\eta}\lesssim 200{\rm\,keV}$
so that the coherent oscillation of the moduli field can be the dark matter and explain the dark matter density.

As mentioned in the introduction,
the oscillating energy density of the moduli field easily exceeds the critical density.
The VEV of the moduli field is expected to be of the order of the Planck scale
during and after the primordial inflation
since it is a singlet field without any symmetry enhanced points.
When the Hubble parameter decreases to the moduli mass scale after the primordial inflation,
the moduli field starts to oscillate and its energy density scales as $a^{-3}$,
where $a$ is the scale factor.
In the case that the reheating after the primordial inflation occurs after the onset of the moduli oscillation,
the ratio of its energy density to the entropy density is estimated as
\begin{equation}
	\frac{\rho_{\eta}}{s_i}\simeq\frac{T_{\rm RH}}{8}\left(\frac{\eta_0}{M_{\rm pl}}\right)^2\simeq1.3\times10^5{\rm\,GeV}\left(\frac{T_{\rm RH}}{10^6{\rm\,GeV}}\right)\left(\frac{\eta_0}{M_{\rm pl}}\right)^2,
	\label{eq:BB_entropy_1}
\end{equation}
where $s_i$ is the entropy density, 
$T_{\rm RH}$ is the reheating temperature 
and $M_{\rm pl}~(\simeq 2.4\times10^{18}{\rm\,GeV})$ is the reduced Planck scale.
$\eta_0$ denotes the initial amplitude of the moduli oscillation,
which is expected to be of the order of $M_{\rm pl}$.
On the other hand,
in the case that the reheating occurs before the onset of the moduli oscillation,
the ratio is given by
\begin{equation}
	\frac{\rho_{\eta}}{s_i}\simeq\frac{1/2m_{\eta}^2\eta_0^2}{2\pi^2/45g_{*}T_{\eta}^3}
	\simeq1.2\times10^6{\rm\,GeV}\left(\frac{m_{\eta}}{200{\rm\,keV}}\right)^{1/2}
	\left(\frac{\eta_0}{M_{\rm pl}}\right)^2, 
	\label{eq:BB_entropy_2}
\end{equation}
where $g_{*}$ is the effective number of degrees of freedom.
$T_{\eta}$ is a temperature at the onset of the moduli oscillation
and is defined as
\begin{equation}
	T_{\eta}\equiv\left(\frac{90}{\pi^2g_*}\right)^{1/4}\sqrt{m_{\eta}M_{\rm pl}}
	\simeq9.7\times10^6{\rm\,GeV}\left(\frac{m_{\eta}}{200{\rm\,keV}}\right)^{1/2}, 
	\label{eq:T_eta}
\end{equation}
where we use $g_*=229$.
The moduli-to-entropy density ratios are conserved until present unless entropy production occurs.
From Eqs.~(\ref{eq:critical_density}), (\ref{eq:BB_entropy_1}) and (\ref{eq:BB_entropy_2}),
one can see that the produced moduli energy density is much larger than the critical density.
This is called ``the cosmological moduli problem".

\subsection{Model of thermal inflation with $Z_4$ symmetry}
\label{subsec:Z_4}

In this subsection,
we explain a model of the thermal inflation with $Z_4$ symmetry
and show that it is excluded because of a late-time decay of ``$R$-axions".

The thermal inflation~\cite{Lyth:1995ka} is an attractive solution to the cosmological moduli problem.
It requires a singlet field $X$ under the standard gauge symmetries, which is called ``a flaton".
We assume that it has a relatively flat potential by imposing a $Z_4$ symmetry\footnote{
Even if we impose $Z_n$ $(n\geq 4)$ symmetry to the superpotential,
the following discussion is almost the same.
}
and has interactions with particles in the thermal bath:
\begin{equation}
	W=\frac{\lambda_{X}}{4M_{\rm pl}}X^4+g_{\xi}X\xi\bar{\xi}+C,
	\label{eq:W_1}
\end{equation}
where $\lambda_X$ and $g_{\xi}$ is dimensionless coupling constants 
and $X$ is the supermultiplet of the flaton field.
$\xi$ and $\bar{\xi}$ are massless gauge charged fields in the thermal bath.
$C$ is a constant term which cancels out the vacuum energy
and is related to the gravitino mass by $|C|\simeq m_{3/2}M_{\rm pl}^2$.
We can ignore higher dimensional terms since the VEV of the flaton is much smaller than the Planck scale.

Motivated by solving the $\mu$ problem in the Higgs sector,
we assign $Z_4$ charge to the minimal SUSY Standard Model (MSSM) particles,
as shown in Table.~\ref{table:Z_4}.
\begin{table}[tb]
\centering
	\begin{tabular}{c|cccccccc}
	supermultiplets&$Q$&$\bar{u}$&$\bar{d}$&$L$&$\bar{e}$&$H_u$&$H_d$&$X$ \\ \hline
	$Z_4$~charge&2&2&0&0&2&0&2&1 
	\end{tabular}
	\caption{Assignment of $Z_4$ charge.}
	\label{table:Z_4}
\end{table}
We assume that the $R$-parity is also conserved in the MSSM sector.
The $Z_4$ symmetry allows the following term~\cite{Kim:1983dt,Stewart:1996ai}:
\begin{equation}
	W_{h}=\frac{\lambda_{\mu}}{M_{\rm pl}}X^2H_uH_d,
	\label{eq:mu-term}
\end{equation}
where $H_u$ and $H_d$ denote up-type and down-type Higgs supermultiplets.
$\lambda_{\mu}$ is a dimensionless coupling.
Although the $\mu$-term is absent due to the $Z_4$ symmetry,
the coupling of Eq.~(\ref{eq:mu-term})
generates an effective $\mu$-term, which is necessary for the electroweak symmetry breaking,
since the flaton acquires the VEV at low energy as we will explain below.

Let us focus on dynamics of the flaton field in the early universe.
At finite temperature, the flaton potential is given by\footnote{
Note that since particles which couple with the flaton acquire mass about $g_{\xi}\left\langle X\right\rangle$,
the flaton acquires the thermal mass term only around the origin of the potential, i.e.,
only when $g_{\xi}\left\langle X\right\rangle\ll T$.
}
\begin{equation}
	V(X)\simeq V_0+(cT^2-m_0^2)|X|^2+\left(\frac{C\lambda_X}{4M_{\rm pl}^3}X^4+{\rm h.c.}\right)+\frac{|\lambda_{X}|^2}{M_{\rm pl}^2}|X|^6,
	\label{eq:original_flaton_potential}
\end{equation}
where we use the same letter $X$ to denote the complex scalar component of the flaton supermutiplet.
$c$ is of the order of the square of the coupling constant $g_{\xi}$.
Here, we add the negative mass term ($-m_0^2<0$) induced from 
gauge-mediated SUSY breaking effects~\cite{Murayama:1992dj},\footnote{
Although the flaton is a singlet field,
the soft SUSY breaking mass term arises 
because it has a coupling with massless gauge charged fields, $\xi$ and $\bar{\xi}$.
}
which is essential for the thermal inflation to work.
When the flaton is near the origin
and $cT^2\gtrsim m_0^2$, 
the flaton field is trapped at the origin.\footnote{
The initial condition is naturally explained by the effects of supergravity.
A positive Hubble induced mass term,
which dominates over the soft mass term in the early universe,
sets the flaton field at the origin.
After that,
it is trapped at the origin by the thermal mass term.
}
Then, 
the vacuum energy $V_0$ becomes to dominate the energy density of the universe
and exponential expansion of the universe, called ``thermal inflation",
occurs.
The thermal inflation lasts until the temperature decreases to the critical value of
$T_c\sim c^{-1/2}m_0$.
At the temperature of $T_c$,
the flaton starts to oscillate around the true minimum given by
\begin{equation}
	\left\langle X\right\rangle\equiv M\simeq\left(\frac{1}{3}\right)^{1/4}\sqrt{\frac{m_0M_{\rm pl}}{|\lambda_X|}},
	\label{eq:M}
\end{equation}
and the thermal inflation ends.
The flaton continues to oscillate until it decays,
which releases huge entropy,
and then the radiation dominated universe is realized.
This is the mechanism of the thermal inflation,
which dilutes the moduli relic by the huge entropy production of the flaton decay.

At low energy,
the flaton field $X$ acquires the VEV and is decomposed as follows:
\begin{equation}
	X=\left(M+\frac{1}{\sqrt{2}}\chi\right)\exp\left(\frac{ia}{\sqrt{2}M}\right),
\end{equation}
where $\chi$ and $a$ are canonically normalized real scalar fields.
The component $\chi$ corresponds to the flaton,
which starts to oscillate after the thermal inflation.
If $C$ was absent in Eq.~(\ref{eq:W_1}),
$X$ would have a $U(1)_R$ symmetry and
the field $a$ would have a flat potential.
We then refer to the scalar field $a$ as ``an $R$-axion",
which obtains small mass from the nonzero constant term $C$
($\simeq m_{3/2}M_{\rm pl}^2$).
The masses of the flaton $\chi$ and the $R$-axion $a$ are calculated as
\begin{equation}
	m_{\chi}^2\simeq4m_0^2,~~~m_{a}^2\simeq\frac{4}{\sqrt{3}}m_0m_{3/2},
\end{equation}
respectively.
Note that $m_{\chi}\gg m_{a}$ since the flaton obtains mass of the order of the SUSY breaking scale
($m_{\chi}\simeq{\cal O}(10^2\mathchar`-10^3){\rm\,GeV}$) and $m_0\gg m_{3/2}$.

Let us consider the decay process of the flaton $\chi$.
The flaton can directly decay into the lightest Higgs boson particles
via the interaction of Eq.~(\ref{eq:mu-term})
when $m_{\chi}\gtrsim 2m_h$,\footnote{
When $m_{\chi}\lesssim 2m_h$,
the flaton decays into the Standard Model particles through one-loop diagrams.
}
where $m_h$ denotes the Higgs boson mass.
The rate of this process is given by~\cite{Asaka:1997rv,Asaka:1999xd}
\begin{equation}
	\Gamma(\chi\to 2h)\simeq\frac{1}{16\pi}\left(\frac{|\lambda_{\mu}|M^2}{M_{\rm pl}m_{\chi}}\right)^4\frac{m_{\chi}^3}{M^2}=\frac{1}{16\pi}\left(\frac{\mu}{m_{\chi}}\right)^4\frac{m_{\chi}^3}{M^2},
	\label{eq:Gamma_2h}
\end{equation}
where $\mu\equiv|\lambda_{\mu}|M^2/M_{\rm pl}$.
Note that the flaton can decay into two higgsinos if it is kinematically allowed,
which may lead to overproduction of the LSP.
We therefore assume that
such a decay is kinematically forbidden,
i.e., $m_{\chi}\lesssim2\mu$.

The flaton can also decay into $R$-axions since $m_{\chi}> m_{a}$.
At the tree level,
the rate of this process is estimated as~\cite{Asaka:1997rv,Asaka:1999xd}
\begin{equation}
	\Gamma(\chi\to2a)\simeq\frac{1}{64\pi}\frac{m_{\chi}^3}{M^2}.
\end{equation}
The produced $R$-axions decay into Standard Model particles through one-loop diagrams,
which is induced by the Yukawa coupling in Eq.~(\ref{eq:W_1}) and the coupling of Eq.~(\ref{eq:mu-term}).
In particular,
$R$-axions decay into photons
and its rate is estimated as~\cite{Asaka:1997rv,Asaka:1999xd}
\begin{equation}
	\Gamma(a\to2\gamma)\simeq\frac{49}{72\pi}\left(\frac{\alpha_{em}}{4\pi}\right)^2\frac{m_a^3}{M^2},
	\label{eq:a_to_SM}
\end{equation}
where $\alpha_{em}$ is the fine-structure constant defined as $\alpha_{em}\equiv e^2/4\pi$.
If loop running particles are charged under $SU(3)_C$ and $m_a\gtrsim 1{\rm\,GeV}$,
the $R$-axion can also decay into gluons.

As seen above,
the flaton decays into both the Standard Model particles 
and $R$-axions
when its decay rate becomes comparable to the Hubble parameter.
The produced Standard Model particles thermalize immediately and reheating occurs.
However, the produced $R$-axions cannot thermalize soon
since they have only one-loop suppressed interactions with particles in thermal bath.
Later,
$R$-axions decay into the Standard Model particles
and reheating occurs again.
From Eq.~(\ref{eq:a_to_SM}),
the reheating temperature when $R$-axions decay, $T^a_{\rm RH}$, is estimated as
\begin{eqnarray}
	T^a_{\rm RH}&\simeq&
	\left(\frac{90}{\pi^2 g_*}\right)^{1/4}\sqrt{\Gamma(a\to2\gamma)M_{\rm pl}} \nonumber \\
	&\simeq&4.5{\rm\,keV}|\lambda_X|^{1/2}\left(\frac{\alpha_{em}}{1/137}\right)
	\left(\frac{m_0}{100{\rm\,GeV}}\right)^{1/4}\left(\frac{m_{3/2}}{200{\rm\,keV}}\right)^{3/4},
\end{eqnarray}
where we use $g_*=3.36$.
It is found that the $R$-axions decay at the temperature much lower than ${\cal O}(10){\rm\,MeV}$.
This implies that it destroys light elements synthesized at the BBN era
and spoils the success of the BBN.

\subsection{Model of thermal inflation with a linear term}
\label{subsec:linear_term}

One way to avoid the late-time decay of $R$-axions
in the $Z_4$ symmetric model
is a prohibition of the flaton decay into $R$-axions,
which can be realized by adding a $Z_4$ symmetry breaking term~\cite{Asaka:1997rv,Asaka:1999xd},
\begin{equation}
	\delta W=\alpha X,
	\label{eq:breaking}
\end{equation}
into the superpotential.
Note that $\alpha$ is a parameter with mass dimension 2.
As we will explain below,
this term leads to degeneracy of masses of the flaton and the $R$-axion,
and the flaton decay into $R$-axions is forbidden kinematically.

Here,
we should comment on another motivation to introduce the $Z_4$ symmetry breaking term.
When the superpotential has the $Z_4$ symmetry as we consider in the previous subsection,
the flaton has four degenerate minima and 
the model is faced with the domain wall problem after the thermal inflation.
However,
the $Z_4$ symmetry breaking term of Eq.~(\ref{eq:breaking}) resolves the degeneracy of the vacua,
which leads to fast collapse of domain walls.
The problem can be solved 
if they collapse before their energy density
dominates that of the universe.
Quantitatively,
the condition to solve the domain wall problem is given by~\cite{Hashiba:1997rp,Stewart:1996ai,Vilenkin:1981zs}
\begin{equation}
	|\alpha|\gtrsim\frac{m_{3/2}^2m_{\chi}M}{M_{\rm pl}^2}.
	\label{eq:domain_wall}
\end{equation}
Hereafter,
we consider the case that $|\alpha|$ is large enough to satisfy Eq.~(\ref{eq:domain_wall}).

The flaton potential is now given by
\begin{eqnarray}
	V(X)&=&V_0-\left(2\frac{\alpha C}{M_{\rm pl}^2}X+{\rm h.c.}\right)-m_0^2|X|^2 \nonumber \\
	&&+\left(\frac{\alpha^*\lambda_{X}}{M_{\rm pl}}X^3+{\rm h.c.}\right)
	+\left(\frac{1}{4}\frac{C\lambda_X}{M_{\rm pl}^3}X^4+{\rm h.c.}\right)
	+\frac{|\lambda_X|^2}{M_{\rm pl}^2}|X|^6.
	\label{eq:modified_potential}
\end{eqnarray}
The negative mass term is induced from the gauge-mediated SUSY breaking effect
and is applicable only for the the flaton VEV smaller than the messenger mass scale.
When the flaton VEV becomes larger than the messenger scale,
the negative soft SUSY breaking term becomes suppressed
at large amplitudes.
In the following,
we focus on the parameter region where the flaton VEV at the true minimum
is larger than the messenger mass scale
and the negative mass term is negligible.
In order to obtain the effective $\mu$-term of the order of the electroweak scale
via the coupling of Eq.~(\ref{eq:mu-term}),
the flaton VEV should be larger than ${\cal O}(10^{10}){\rm\,GeV}$.
In this parameter region,
the trilinear term dominates over
both the linear term and the quartic term at the true minimum.
The flaton VEV at present and the vacuum energy at the origin are thus estimated as
\begin{equation}
	\left\langle X\right\rangle\equiv 
	M\simeq\left(\frac{|\alpha| M_{\rm pl}}{|\lambda_X|}\right)^{1/3},~~~V_0\simeq |\alpha|^2,
	\label{eq:M_alpha}
\end{equation}
where $V_0$ is determined by requiring that the vacuum energy vanishes at the true minimum.
The masses of the flaton and the $R$-axion are now almost the same and are given by
\begin{equation}
	m_{\chi}\simeq m_a\simeq3\left(\frac{|\alpha|^2|\lambda_X|}{M_{\rm pl}}\right)^{1/3}.
\end{equation}
Note that the flaton decay into $R$-axions is forbidden since their masses are degenerate.

Taking the thermal effects into account,
the flaton potential around the origin is rewritten by
\begin{equation}
	V(X)\simeq (cT^2-m_0^2)\left|X-\frac{2\alpha^* C}{(cT^2-m_0^2)M_{\rm pl}^2}\right|^2
	+V_0-\frac{4|\alpha|^2C^2}{(cT^2-m_0^2)M_{\rm pl}^4},
\end{equation}
and the local minimum before and during the thermal inflation is given by 
\begin{equation}
	\left\langle |X|\right\rangle\simeq\frac{2|\alpha| m_{3/2}}{cT^2-m_0^2}.
	\label{eq:local_minimum}
\end{equation}
Then,
one can find that the flaton does not sit at the origin due to the $Z_4$ symmetry breaking term $\alpha$.
This implies that the thermal inflation ends
when the VEV of $X$ becomes so large that it decouples from the thermal plasma, i.e.,
when the temperature decreases to $T_c\simeq c^{1/2}\left\langle |X|\right\rangle$.
Although the vacuum energy during the thermal inflation also deviates from $V_0$,
the deviation is estimated as $\delta V/V\simeq m_{3/2}^2/cT^2$
and is negligible.

We turn to estimate how much entropy is released by the flaton decay after the thermal inflation.
We define a dilution factor
as the ratio of the initial to the final comoving entropy density.
It is calculated as~\cite{Asaka:1997rv,Asaka:1999xd}
\begin{equation}
	\Delta\equiv\frac{s_fa_f^3}{s_ia_i^3}=1+\frac{4}{3}\frac{V_0}{2\pi^2/45g_*T_c^3T^{\chi}_{\rm RH}},
	\label{eq:dilution_factor_1}
\end{equation}
where $s_f$ represents the entropy density after the thermal inflation
and $T^{\chi}_{\rm RH}$ denotes the reheating temperature when the flaton decays.
$T^{\chi}_{\rm RH}$ is calculated from its decay rate and has to be higher than ${\cal O}(10){\rm\,MeV}$.

From the dilution factor,
one can estimate the moduli density parameter.
Hereafter,
we refer to the moduli produced before the thermal inflation as ``Big-Bang moduli".
Using Eq.~(\ref{eq:BB_entropy_1}),
the density parameter of the Big-Bang moduli is given by
\begin{eqnarray}
	\Omega_{\eta,{\rm BB}}h^2&=&\frac{1}{\Delta}\frac{\rho_{\eta,{\rm BB}}}{s_i}\frac{s_0h^2}{\rho_{\rm cr}}
	\nonumber \\
	&\simeq&0.026\left(\frac{T_{\rm RH}}{10^6{\rm\,GeV}}\right)\left(\frac{V_0}{10^{24}{\rm\,GeV}^4}\right)^{-1}\left(\frac{T^{\chi}_{\rm RH}}{10{\rm\,MeV}}\right)
	\left(\frac{T_c}{1{\rm\,TeV}}\right)^3\left(\frac{\eta_0}{M_{\rm pl}}\right)^2, 
	\label{eq:BigBang_moduli}
\end{eqnarray}
where the subscripts ``BB" represent the Big-Bang moduli.
If the reheating of the primordial inflation occurs before the moduli oscillation,
one can obtain the density parameter using Eq.~(\ref{eq:BB_entropy_2}).
It is found that
the abundance of the Big-Bang moduli density can be sufficiently diluted
by the thermal inflation.

Unfortunately,
the thermal inflation gives a secondary source of the moduli.
We call the secondary produced moduli as ``thermal inflation moduli".
The potential of the moduli field during the thermal inflation is given by
\begin{eqnarray}
	V(\eta)&=&\frac{1}{2}m_{\eta}^2\eta^2+\frac{c_H}{2}H^2(\eta-\eta_0)^2 \nonumber \\
	&\simeq&\frac{1}{2}(m_{\eta}^2+c_HH^2)\left(\eta-\frac{c_HH^2}{m_{\eta}^2+c_HH^2}\eta_0\right)^2+\dots,
	\label{eq:V_thermal_moduli}
\end{eqnarray}
where $c_H$ is a coefficient of the Hubble induced mass term,
which comes from the finite vacuum energy during the thermal inflation.
$\eta_0$ is a minimum determined by the Hubble induced mass term
and is expected to be of the order of the Planck scale.
Equation (\ref{eq:V_thermal_moduli}) clarifies that
the moduli field sits at the minimum depending on the Hubble induced mass term
during the thermal inflation.
After that,
it starts to roll down to the true minimum with the amplitude of $\eta\simeq c_H\left(H_{\rm th}/m_{\eta}\right)^2\eta_0$,
where $H_{\rm th}$ is the Hubble parameter at the end of the thermal inflation.
The ratio of the moduli to the entropy density at present is therefore given by
\begin{equation}
	\frac{\rho_{\eta,{\rm TH}}}{s_f}=\frac{\rho_{\eta,{\rm TH}}}{4V_0/3T^{\chi}_{\rm RH}}=\frac{c_H^2T^{\chi}_{\rm RH}V_0\eta_0^2}{24m_{\eta}^2M_{\rm pl}^4}.
\end{equation}
The density parameter of the thermal inflation moduli is estimated as
\begin{equation}
	\Omega_{\eta,{\rm TH}}h^2=\frac{\rho_{\eta,{\rm TH}}}{s_f}\frac{s_0h^2}{\rho_{\rm cr}} 
	\simeq 0.48 c_H^2\left(\frac{V_0}{10^{24}{\rm\,GeV}^4}\right)\left(\frac{m_{3/2}}{200{\rm\,keV}}\right)^{-2}\left(\frac{T^{\chi}_{\rm RH}}{10{\rm\,MeV}}\right)\left(\frac{\eta_0}{M_{\rm pl}}\right)^2.
	\label{eq:thermal_moduli}
\end{equation}
Note that the thermal inflation moduli is produced more abundantly
as the vacuum energy $V_0$ becomes higher.
This is contrary to the Big-Bang moduli.

From Eqs.~(\ref{eq:BigBang_moduli}) and (\ref{eq:thermal_moduli}),
it is found that $\Omega_{\eta,{\rm BB}}h^2+\Omega_{\eta,{\rm TH}}h^2\lesssim {\cal O}(0.1)$ 
for $V_0\simeq{\cal O}(10^{24}){\rm\,GeV}^4$,
$T_c\simeq{\cal O}(1){\rm\,TeV}$ 
and $T^{\chi}_{\rm\,RH}\simeq{\cal O}(10){\rm\,MeV}$,
and that the thermal inflation can solve the moduli problem.
The relic of the moduli could be the cold dark matter.

\section{Affleck-Dine baryogengesis and Q-ball}
\label{sec:AD_Q-ball}

As mentioned in the introduction,
the thermal inflation dilutes not only the moduli density 
but also the baryon asymmetry which may be produced beforehand.
Therefore,
some viable mechanisms to produce sufficiently large amount of baryon asymmetry are needed
for a cosmological consistent scenario.
The Affleck-Dine baryogenesis~\cite{Affleck:1984fy,Dine:1995kz} is a promising candidate
in the framework of SUSY.
In the Affleck-Dine mechanism,
flat directions rotating in the complex plane produce baryon number and
they decay into the Standard Model particles in the early universe.
In gauge-mediated SUSY breaking models, however,
the Affleck-Dine field
forms non-topological solitons called ``Q-balls"
and the produced baryon charge is confined into Q-balls~\cite{Coleman:1985ki,Kusenko:1997si,Enqvist:1998en,Kasuya:1999wu}.
We briefly review the Affleck-Dine baryogenesis and Q-ball properties in this section.

\subsection{Affleck-Dine baryogenesis}
\label{subsec:AD}

Supersymmetry predicts
a lot of flat directions which have vanishing potentials
in renormalizable levels.
In particular,
the MSSM contains flat directions carrying baryon (and/or lepton) charge.
They are referred to as ``Affleck-Dine (AD) fields".
One example of them is the $\bar{u}\bar{d}\bar{d}$ flat direction.
The $\bar{u}\bar{d}\bar{d}$ flat direction is a linear combination of
right-handed squark fields with the same amplitude,
for example,
$\tilde{\bar{u}}^{R}_1=\frac{1}{\sqrt{3}}\Phi$,
$\tilde{\bar{d}}^{G}_1=\frac{1}{\sqrt{3}}\Phi$ and
$\tilde{\bar{d}}^{B}_2=\frac{1}{\sqrt{3}}\Phi$,
where $\tilde{\bar{u}}$ and $\tilde{\bar{d}}$ are 
the up-type and down-type right-handed squarks,
respectively.
The superscripts and the subscripts show
color and family indices,
respectively.
In this direction,
$D$-term potentials indeed vanish and
it also has no renormalizable terms in $F$-terms if the $R$-parity is conserved.
In the following,
we explain the Affleck-Dine mechanism without restricting to the $\bar{u}\bar{d}\bar{d}$ flat direction.

The flat directions acquire the potential
from the SUSY breaking effects
and the scalar potential is expressed as
\begin{equation}
	V_{\rm SB}(\Phi)=V_{\rm gauge}+V_{\rm grav}=M_F^4\left(\log\frac{|\Phi|^2}{M_{\rm mess}^2}\right)^2
	+m_{3/2}^2\left(1+K\log\frac{|\Phi|^2}{M_*^2}\right)|\Phi|^2,
	\label{eq:AD_potential}
\end{equation}
where $\Phi$ denotes the AD field.
The first term comes from gauge mediation effects~\cite{de Gouvea:1997tn,Kusenko:1997si}
and is applicable only for $|\Phi|\gg M_{\rm mess}$,\footnote{
For $|\Phi|\ll M_{\rm mess}$,
the first term in Eq.~(\ref{eq:AD_potential}) is replaced by a soft SUSY breaking mass term,
$m_{\rm SUSY}^2|\Phi|^2$,
where $m_{\rm SUSY}\sim{\cal O}(0.1\mathchar`-1){\rm\,TeV}$.
On the other hand,
it becomes logarithmic for $|\Phi|\gg M_{\rm mess}$
since a large amplitude of the AD field leads to
suppression of the transmission of SUSY breaking effects~\cite{de Gouvea:1997tn}.
}
where $M_{\rm mess}$ is the messenger mass scale.
$M_{F}$ is related to SUSY breaking scale as follows:
\begin{equation}
	M_F\simeq\frac{g^{1/2}}{4\pi}\sqrt{kF}
\end{equation}
where $g$ generically represents gauge couplings of the Standard Model
and $F$ denotes the SUSY breaking $F$-term.
The parameter $k$ is determined from couplings
between the SUSY breaking sector and the messenger sector and satisfies $k\leq1$.
The SUSY breaking scale is constrained from below 
by the relatively heavy Higgs boson mass at around $126{\rm\,GeV}$,
which acquires the radiative corrections from the stop mass.
Since the masses of scalar particles are proportional to the parameter $\Lambda\equiv kF/M_{\rm mess}$,
we obtain a lower bound of $\Lambda$ as follows~\cite{Hamaguchi:2014sea,Kasuya:2015uka}:
\begin{equation}
	\Lambda\equiv\frac{kF}{M_{\rm mess}}\gtrsim5\times10^5{\rm\,GeV}.
\end{equation}
In combination with $M_{\rm mess}^2\gtrsim kF$
which is necessary to make masses of messenger scalars positive,
$\sqrt{kF}\gtrsim\Lambda\gtrsim 5\times10^5{\rm\,GeV}$
must be realized.
Therefore,
$M_F$ is restricted from below,
\begin{equation}
	M_F\gtrsim4\times10^4g^{1/2}{\rm\,GeV}.
	\label{eq:M_F}
\end{equation}
Since the SUSY breaking scale $F$ is related with the gravitino mass as
$F\simeq \sqrt{3}m_{3/2}M_{\rm pl}$,
$M_F$ is restricted from the above as follows:
\begin{equation}
	M_F\lesssim2\times10^6{\rm\,GeV}g^{1/2}\left(\frac{m_{3/2}}{200{\rm\,keV}}\right)^{1/2}. 
\end{equation}
The second term of Eq.~(\ref{eq:AD_potential}) comes from the gravitational mediation effects
including one-loop corrections~\cite{Enqvist:1997si}.
$M_*$ is a renormalization scale.
The parameter $K$ comes from 1-loop effects and its absolute value
is typically in the range of 0.01 to 0.1.
If gaugino contribution to 1-loop effects is larger than that of the Yukawa interactions,
the sign of $K$ is negative, and vice versa.
Notice that the gravitational mediation effects dominate over the gauge mediation effects
above a threshold value $\phi_{\rm eq}$ which is given by 
\begin{equation}
	\phi_{\rm eq}\simeq3.2\times10^{14}{\rm\,GeV}
	\left(\frac{m_{3/2}}{200{\rm\,keV}}\right)^{-1}
	\left(\frac{M_F}{4\times10^{4}{\rm\,GeV}}\right)^2,
	\label{eq:phi_eq_value}
\end{equation}
for $M_{\rm mess}\simeq5\times10^5{\rm\,GeV}$.

There also exists thermal effects on the effective potential.
When $f|\Phi|\ll T$,where $f$ is a coupling constant between thermal particles and the AD field,
the thermal mass term arises and is given by
\begin{equation}
V^{(1)}_{\rm th}\sim f^2T^2|\Phi|^2.
\end{equation}
A thermal effect via 2-loop diagrams
also contributes to the effective potential for $f|\Phi|\gg T$ as follows:
\begin{equation}
	V^{(2)}_{\rm th}\sim\alpha_g^2 T^4\log\frac{|\Phi|^2}{T^2},
	\label{eq:thermal_log}
\end{equation}
where $\alpha_g=g^2/4\pi$~\cite{Anisimov:2000wx,Fujii:2001zr}.

The flat directions are also lifted by non-renormalizable superpotential
and we generically write it as
\begin{equation}
	W_{\rm NR}=\frac{\lambda_{\Phi}\Phi^n}{nM_{\rm pl}^{n-3}},
	\label{eq:NR}
\end{equation}
where $\lambda_{\Phi}$ is a coupling constant 
and $n~(\geq 4)$ is an integer.
For example,
in the case of $\bar{u}\bar{d}\bar{d}$ flat direction,
$n=6$.
The superpotential of Eq.~(\ref{eq:NR}) leads to the non-renormalizable terms in the AD field potential as
\begin{equation}
	V_{\rm NR}(\Phi)=\lambda_{\Phi}a_{\lambda}m_{3/2}\frac{\Phi^n}{M_{\rm pl}^{n-3}}+{\rm h.c.}+|\lambda_{\Phi}|^2\frac{|\Phi|^{2n-2}}{M_{\rm pl}^{2n-6}}.
	\label{eq:AD_potential_2}
\end{equation}
Here,
we introduce a dimensionless parameter $a_{\lambda}$ of ${\cal O}(1)$,
which depends on higher dimensional K\"ahler potentials.
The terms proportional to $a_{\lambda}$ are induced from the gravitational effects
and called ``$A$-terms".

In addition,
the AD field acquires Hubble induced terms in the early universe.
The vacuum energy causing the expansion of the universe largely
breaks the supersymmetry
and lifts the potential of the flat direction.
Since the vacuum energy is related to the Hubble parameter,
these additional terms are given by
\begin{equation}
	V_H=-\bar{c}_HH^2|\Phi|^2+\lambda_{\Phi}a_HH\frac{\Phi^n}{M_{\rm pl}^{n-3}}+{\rm h.c.},
	\label{eq:AD_potential_Hubble}
\end{equation}
where $\bar{c}_H$ and $a_H$ are dimensionless parameters of ${\cal O}$(1).
Hereafter,
we consider the case of $\bar{c}_H>0$
to realize the Affleck-Dine baryogenesis.

Next, let us explain dynamics of the AD field in the early universe.
Its potential is given by Eqs.~(\ref{eq:AD_potential}), (\ref{eq:AD_potential_2}) and (\ref{eq:AD_potential_Hubble}).
During the primordial inflation,
the AD field sits down at the minimum determined by 
the negative Hubble induced mass term 
and the non-renormalizable terms in Eq.~(\ref{eq:AD_potential_2}).
Its VEV is estimated as
\begin{equation}
	|\Phi|\sim\left(HM_{\rm pl}^{n-3}/|\lambda_{\Phi}|\right)^{1/(n-2)},
	\label{eq:AD_VEV}
\end{equation}
and the phase of the AD field is determined by the Hubble induce $A$-term.
After the inflation end,
the energy density of the universe is dominated by 
the oscillating inflaton
and the Hubble parameter decreases.
When $H\sim V'({\Phi})/\Phi$,
the AD field starts to roll down to the origin of the potential.
At this time,
the phase of the AD field also starts to rotate in the complex plane
producing baryon number, $n_B\simeq{\rm Im}[\Phi^*\dot{\Phi}]$.
The produced baryon number density at the
onset of the oscillation of the AD field is roughly estimated as
\begin{eqnarray}
	\left.n_B\right|_{\rm AD~osc}&\sim& 
	H^{-1}{\rm Im}\left.\left[\frac{\partial V}{\partial \Phi}\Phi\right]\right|_{\rm AD~osc}
	\sim \frac{m_{3/2}|\lambda_{\Phi}||\Phi_{\rm osc}|^n}{H_{\rm osc}M_{\rm pl}^{n-3}}
	\sim \left(\frac{m_{3/2}}{H_{\rm osc}}\right)H_{\rm osc}|\Phi_{\rm osc}|^2 \nonumber \\
	&\equiv&\epsilon H_{\rm osc}|\Phi_{\rm osc}|^2,
\end{eqnarray}
where we use Eq.~(\ref{eq:AD_VEV})
and define the ellipticity parameter $\epsilon\equiv m_{3/2}/H_{\rm osc}$.
As the oscillation amplitude of the AD field decreases due to the Hubble friction,
higher order terms become irrelevant.
Therefore,
the baryon number violating operators,
the $A$-terms,
produce baryon number only at the onset of the oscillation.
After that,
the comoving density of the baryon number is conserved.
The produced baryon number is
finally converted into the Standard Model particles
by the decay of the AD field
and the present baryon asymmetry can be explained.
This is the mechanism of the Affleck-Dine baryogenesis.

\subsection{Q-ball}
\label{subsec:Q-ball}

It is known that the AD field oscillating in the potential of Eq.~(\ref{eq:AD_potential})
feels spatial instabilities and forms into non-topological solitons, Q-balls~\cite{Coleman:1985ki}.
Let us explain a condition that Q-balls are formed after the Affleck-Dine baryognesis.
The configuration of the Q-ball
is determined by the condition of minimizing the energy with conserved baryon charge,
where the energy and the baryon charge are given by
\begin{equation}
	E=\int d^3x \left(|\dot{\Phi}|^2+|\nabla\Phi|^2+V(|\Phi|)\right),~~~Q=2\int d^3x {\rm Im}\left[\Phi^*\dot{\Phi}\right],
	\label{eq:Q-ball_energy_E}
\end{equation}
respectively.
Here,
we assume that the AD field carries an unit of baryon charge for simplicity.
The scalar field configuration is obtained by minimizing
\begin{equation}
	E_{\omega}\equiv E+\omega\left(Q-2\int d^3x {\rm Im}\left[\Phi^*\dot{\Phi}\right]\right),
\end{equation}
where $\omega$ is a Lagrange multiplier.
$E_{\omega}$ is rewritten by
\begin{equation}
	E_{\omega}=\int d^3x\left[|\dot{\Phi}-i\omega\Phi|^2-\omega^2|\Phi|^2+|\nabla\Phi|^2+V(|\Phi|)\right]+\omega Q.
\end{equation}
The time dependence of $\Phi$ is determined as $\Phi(\bold{x},t)=\varphi(r) e^{i\omega t}/\sqrt{2}$
from the first term
in order to minimize $E_{\omega}$.
We assume that the stable field configuration is spherically symmetric,
which leads to the following equation of the field configuration:
\begin{equation}
	\frac{\partial^2}{\partial r^2}\varphi+\frac{2}{r}\frac{\partial}{\partial r}\varphi
	+\omega^2\varphi-\frac{\partial}{\partial \varphi}V(\varphi)=0.
	\label{eq:field_configuration_equation}
\end{equation}
The boundary condition is $\varphi'(0)=0$ and $\varphi(\infty)=0$
in order to obtain a smooth and local configuration.
The solution with the boundary condition exits for
\begin{equation}
	\omega_0^2\equiv{\rm min}\left[\frac{2V(\varphi)}{\varphi^2}\right]_{\varphi=\varphi_0\neq0}<\frac{\partial^2 V(0)}{\partial \varphi^2}.
	\label{eq:Q-ball_condition}
\end{equation}
This inequality requires that there exists a field value 
where the potential is flatter than the quadratic potential.
This condition is satisfied if the potential of the AD field
is dominated by the first term of Eq.~(\ref{eq:AD_potential}).
Even if the second term dominates over the first one,
the condition of Eq.~(\ref{eq:Q-ball_condition}) is satisfied for the case of $K<0$.
Numerical simulations have shown that 
AD fields which oscillate in the potential satisfying Eq.~(\ref{eq:Q-ball_condition})
actually form into Q-balls~\cite{Kasuya:1999wu,Kasuya:2000sc,Kasuya:2001hg}.
It has also been shown that 
(almost) all baryon number is confined into Q-balls.

The profile of the Q-ball depends on the potential of the AD field.
When the gauge-mediation effect dominates over the gravity-mediation one,
formed Q-balls are referred to as ``gauge-mediation type Q-balls".
The field configuration of the gauge-mediation type Q-ball is determined by solving Eq.~(\ref{eq:field_configuration_equation}) and
is approximately given by~\cite{Dvali:1997qv}
\begin{eqnarray}
	\Phi(r)\simeq \frac{e^{i\omega t}}{\sqrt{2}}\times\left\{
	\begin{array}{ll}
		\varphi_0\frac{\sin\omega r}{\omega r} & {\rm for~}r<R\equiv\pi/\omega \\
		0 & {\rm for~}r>R \\
	\end{array}
	\right.,
	\label{eq:GMSB_config}
\end{eqnarray}
where $\omega$ and $\varphi_0$ are given by
\begin{equation}
	\omega\simeq \sqrt{2}\pi M_FQ^{-1/4},~~~\varphi_0\simeq M_FQ^{1/4}.
	\label{eq:GMSB_profile}
\end{equation}
The energy of the Q-ball is calculated from Eq.~(\ref{eq:Q-ball_energy_E}) and is estimated as
\begin{equation}
	E\simeq \frac{4\sqrt{2}\pi}{3}M_FQ^{3/4}.
	\label{eq:GMSB_energy}
\end{equation}
One can find that the Q-ball energy per unit charge ($\simeq dE/dQ$) is smaller for larger $Q$.
When it is smaller than the proton mass,
i.e.,
$dE/dQ\sim M_FQ^{-1/4}<1{\rm\,GeV}$,
Q-balls cannot decay into nucleons.
The charge of the Q-ball can be determined by numerical simulations
and is given by~\cite{Kasuya:2001hg}
\begin{equation}
	Q\sim \beta\left(\frac{\phi_{\rm osc}}{M_F}\right)^4,
	\label{eq:Q_charge}
\end{equation}
where $\beta$ is a numerical coefficient and is determined as $\beta\simeq6\times10^{-4}$.\footnote{
Precisely speaking,
the numerical coefficient $\beta$ depends on the orbit of the AD field in the complex plane.
$\beta\simeq 6\times10^{-4}$ for a circular orbit ($\epsilon=1$),
while $\beta\simeq 6\times 10^{-5}$ for an oblate orbit ($\epsilon\lesssim 0.1$).
Hereafter,
we use $\beta\simeq 6\times10^{-4}$ for simplicity
since our results do not change significantly.
}
Hereafter,
we use $\phi$ as the amplitude of the AD field and
$\phi_{\rm osc}$ denotes the amplitude at the onset of the oscillation,
$|\Phi_{\rm osc}|$.
Note that $\phi_{\rm osc}$ should be smaller than $\phi_{\rm eq}$ to use Eq.~(\ref{eq:Q_charge}).

In the case of $\phi_{\rm osc}\gtrsim \phi_{\rm eq}$,
the gravity mediation effect dominates over the gauge mediation one and
one can consider two types of scenarios:
$K>0$ and $K<0$.
When $K<0$,
the second term of Eq.~(\ref{eq:AD_potential}) satisfies the condition for Q-ball formation.
Q-balls formed in this potential are referred to as ``new type Q-balls"~\cite{Kasuya:2000sc}.
The field configuration is approximately given by a Gaussian function~\cite{Enqvist:1997si}:
\begin{equation}
	\Phi(r)\simeq \frac{1}{\sqrt{2}}\varphi_0e^{-r^2/2R^2}e^{i\omega t},
\end{equation}
where $R$, $\omega$ and $\varphi_0$ are given by
\begin{equation}
	R\simeq\frac{1}{|K|^{1/2}m_{3/2}},~~~\omega\simeq m_{3/2},~~~\varphi_0\simeq \left(\frac{|K|}{\pi}\right)^{3/4}m_{3/2}Q^{1/2},
\end{equation}
respectively.
The energy of the Q-ball is given by
\begin{equation}
	E\simeq m_{3/2}Q.
\end{equation}
This type of the Q-balls is stable against the decay into nucleons
since $dE/dQ\simeq m_{3/2}<1{\rm\,GeV}$ in gauge-mediated SUSY breaking models.
The charge of the new type Q-ball is given by~\cite{Kasuya:2000wx,Hiramatsu:2010dx}
\begin{equation}
	Q\sim\tilde{\beta}\left(\frac{\phi_{\rm osc}}{m_{3/2}}\right)^2
\end{equation}
where $\tilde{\beta}\simeq 2\times 10^{-2}$.
When $K>0$,
on the other hand,
the condition of Eq.~(\ref{eq:Q-ball_condition}) is not satisfied and
Q-balls are not formed.
In this case,
the oscillation of the AD field remains homogeneous and 
its amplitude decreases as $\phi\propto a^{-3/2}$
after it starts to oscillate.
However,
Q-balls are formed
when its amplitude decreases to $\phi_{\rm eq}$
and the potential of the AD field 
becomes dominated by the gauge mediation effect.
This type of Q-balls is referred to as ``delayed type Q-balls"~\cite{Kasuya:2001hg}.
The profile and properties of the Q-ball are the same as those of the gauge-mediation type Q-ball
(see Eqs.~(\ref{eq:GMSB_config}), (\ref{eq:GMSB_profile}) and (\ref{eq:GMSB_energy})),
while the charge of the delayed type Q-ball is given by
\begin{equation}
	Q\sim\beta\left(\frac{\phi_{\rm eq}}{M_F}\right)^4.
\end{equation}

The thermal logarithmic potential of Eq.~(\ref{eq:thermal_log}) also satisfies the condition for Q-ball formation.
If the AD field starts to oscillate by the thermal logarithmic potential,
one can obtain the charge of the Q-ball
by replacing $M_F$ with $T_*$ in Eq.~(\ref{eq:Q_charge}),
where $T_*$ is the temperature at Q-ball formation.
The profile and properties become the same as those of the gauge-mediation type Q-balls
when the temperature decreases sufficiently.

\section{Moduli problem and Baryon asymmetry with Q-ball formation}
\label{sec:moduli_Affleck}

In gauge-mediated SUSY breaking models,
Q-balls are formed during the oscillation of the AD field.
Almost all baryon number is absorbed into Q-balls
even if huge baryon asymmetry is produced by the Affleck-Dine baryogenesis.
If Q-balls are stable,
baryon charge is released from Q-balls via evaporation~\cite{Laine:1998rg}.
In the case of unstable Q-balls,
the decay of Q-balls can release baryon charge.
However,
it must occur before the BBN epoch in order to explain the baryon asymmetry.
In this section,
we show that the baryon number production is incompatible with the moduli dilution
in both cases updating the analysis in~\cite{Kasuya:2001tp}.

\subsection{Stable Q-ball}
\label{subsec:stable}

First,
we consider stable Q-balls and
briefly review that it is difficult to explain the baryon asymmetry in this case.
The Q-ball is stable against the decay into nucleons
when $dE/dQ<1{\rm\,GeV}$.
At finite temperature,
however,
baryon charge can evaporate from Q-ball surface
since the free energy of the AD field in the thermal plasma 
is smaller than that of Q-balls.
The total evaporated charge of gauge-mediation type Q-balls is given by~\cite{Kasuya:2001hg,Kasuya:2014ofa}
\begin{equation}
	\Delta Q\sim10^{16}\left(\frac{m_{\rm SUSY}}{1{\rm\,TeV}}\right)^{-2/3}
	\left(\frac{M_F}{4\times10^4{\rm\,GeV}}\right)^{-1/3}Q^{1/12},
\end{equation}
where $m_{\rm SUSY}$ denotes a soft SUSY breaking mass scale.
In the case of new type Q-balls,
it is given by~\cite{Kasuya:2000sc,Kasuya:2014ofa}
\begin{equation}
	\Delta Q\sim10^{20}\left(\frac{|K|}{0.01}\right)^{-2/3}\left(\frac{m_{3/2}}{200{\rm\,keV}}\right)^{-1/3}
	\left(\frac{m_{\rm SUSY}}{1{\rm\,TeV}}\right)^{-2/3}.
\end{equation}
Since $\Delta Q/Q$ becomes smaller as $Q$ increases,
it is more difficult to extract baryon charge from Q-balls with larger baryon number.
In order to produce huge baryon number,
$\phi_{\rm osc}$ should be large,
which increases the baryon charge of the Q-ball.
Then,
the released baryon number generally becomes smaller
as $\phi_{\rm osc}$ increases.
In the case of the delayed type Q-ball,
however,
the baryon charge of Q-balls does not increase
even though $\phi_{\rm osc}$ becomes much larger than $\phi_{\rm eq}$.
Hence,
the delayed type Q-ball seems to be able to most effectively provide baryon charge outside Q-balls.
We thus focus on the delayed type Q-ball in the following.
For the delayed type Q-ball,
$\Delta Q/Q$ is estimated as
\begin{equation}
	\frac{\Delta Q}{Q}\sim 4\times 10^{-18}\left(\frac{m_{\rm SUSY}}{1{\rm\,TeV}}\right)^{-2/3}
	\left(\frac{M_F}{4\times10^4{\rm\,GeV}}\right)^{-4}\left(\frac{m_{3/2}}{200{\rm\,keV}}\right)^{11/3},
	\label{eq:delta_Q/Q}
\end{equation}
where we use Eq.~(\ref{eq:phi_eq_value}).

The finally provided baryon asymmetry is calculated from
\begin{equation}
	Y_B\equiv\frac{n_B}{s}=\frac{\tilde{n}_{B}}{s}\frac{\Delta Q}{Q}
	=\left.\frac{\tilde{n}_{B}}{\rho_{\eta,{\rm BB}}}\right|_{\eta{\rm~osc}}
	\left.\frac{\rho_{\eta,{\rm BB}}}{s}\right|_{0}\frac{\Delta Q}{Q},
	\label{eq:baryon_asymmetry_1}
\end{equation}
where $\tilde{n}_{B}/s$ is the ratio estimated without considering Q-ball formation.
When $\phi_{\rm osc}\gtrsim\phi_{\rm eq}$,
the ratio of the baryon number to the Big-Bang moduli energy density 
is estimated as
\begin{eqnarray}
	\left.\frac{\tilde{n}_B}{\rho_{\eta,{\rm BB}}}\right|_{\eta{\rm~osc}}
	=\left.\frac{\tilde{n}_B}{3M_{\rm pl}^2H^2}\right|_{\rm AD~osc}
	\left.\frac{3M_{\rm pl}^2H^2}{\rho_{\eta,{\rm BB}}}\right|_{\eta{\rm~osc}}
	&\simeq&\frac{2m_{3/2}}{H_{\rm osc}^2}\left(\frac{\phi_{\rm osc}}{M_{\rm pl}}\right)^2
	\left(\frac{\eta_0}{M_{\rm pl}}\right)^{-2} 
	\label{eq:n_B/rho_eta} \\
	&\simeq&\frac{2}{m_{3/2}}\left(\frac{\phi_{\rm osc}}{M_{\rm pl}}\right)^2
	\left(\frac{\eta_0}{M_{\rm pl}}\right)^{-2},
	\label{eq:n/rho}
\end{eqnarray}
where we assume $H_{\rm osc}\simeq m_{3/2}$ in the second line,
which is the case of new or delayed type Q-balls.
Here, 
we assume that the total energy density of the universe scales as $a^{-3}$
between the eras when the AD field starts to oscillate and when the moduli field starts to oscillate,
since the moduli field starts to oscillate before the reheating.\footnote{
$V_{\rm gauge}\gtrsim V^{(2)}_{\rm th}$ and $V_{\rm gauge}\gtrsim V_{\rm grav}$
lead to $\phi_{\rm osc}/M_{\rm pl}\gtrsim(T_{\rm RH}/M_F)^2$
and $\phi_{\rm osc}/M_{\rm pl}\lesssim M_F^2/m_{3/2}M_{\rm pl}$.
Combining these relations,
$T_{\rm RH}\lesssim M_F^2/\sqrt{m_{3/2}M_{\rm pl}}$
should be realized.
Since $M_F^2\lesssim m_{3/2}M_{\rm pl}$,
the reheating temperature is estimated as $T_{\rm RH}\lesssim T_{\eta}$,
which implies the reheating occurs after the moduli oscillation.
}
From Eqs.~(\ref{eq:delta_Q/Q}) and (\ref{eq:n/rho}),
the baryon asymmetry is expressed as 
\begin{eqnarray}
	Y_B&\sim&2\times10^{-23}\left(\frac{\Omega_{\eta,{\rm BB}}h^2}{0.12}\right)
	\left(\frac{\eta_0}{M_{\rm pl}}\right)^{-2}
	\left(\frac{m_{3/2}}{200{\rm\,keV}}\right)^{8/3} \nonumber \\
	&&\times\left(\frac{m_{\rm SUSY}}{1{\rm\,TeV}}\right)^{-2/3}\left(\frac{M_F}{4\times10^4{\rm\,GeV}}\right)^{-4}
	\left(\frac{\phi_{\rm osc}}{M_{\rm pl}}\right)^2,
\end{eqnarray}
where we use the Big-Bang moduli density after the entropy dilution due to the thermal inflation
and assume that the abundance of the thermal inflation moduli is negligible.
Here,
we assume that the thermal effects on the effective potential is negligible.
The stability condition,
$dE/dQ< 1{\rm\,GeV}$,
leads to the upper bound on the gravitino mass,
$m_{3/2}\lesssim 1.4{\rm\,GeV}$.
One can find that the estimated baryon asymmetry is too small to explain the present one,
$Y_{B}\simeq 8\times10^{-11}$,
even if $\phi_{\rm osc}\sim M_{\rm pl}$
and $M_F$ taken as the lower bound of Eq.~(\ref{eq:M_F}),
which puts the more severe upper bound of the produced baryon number than~\cite{Kasuya:2001tp}.
Other scenarios,
including ones where Q-balls are formed by the thermal logarithmic potential,
cannot explain the present baryon asymmetry, either~\cite{Kasuya:2001tp}.

\subsection{Unstable Q-ball}
\label{subsec:unstable}

We show that the situation does not become improved even in the case of the unstable Q-ball.
It can decay into nucleons 
when $dE/dQ>1{\rm\,GeV}$
and release baryon charge from its surface.
The emission rate is determined by the Pauli blocking effect on its surface
and is given by~\cite{Cohen:1986ct,Kawasaki:2012gk,Harigaya:2014tla}
\begin{equation}
	\left|\frac{dQ}{dt}\right|\simeq\frac{(2\omega)^3A}{96\pi^2},
\end{equation}
where $A$ is the surface area of the Q-ball.
In order to avoid destroying light elements due to the decay products of Q-balls,
the decay temperature is constrained as follows:
\begin{equation}
	T_{\rm dec}\equiv\sqrt{\frac{1}{Q}\frac{dQ}{dt}M_{\rm pl}}
	\gtrsim {\cal O}(10){\rm\,MeV}.
	\label{eq:Q-ball_decay}
\end{equation}

In gauge-mediated SUSY breaking models,
unstable Q-balls correspond to ``gauge-mediation type Q-balls",
where the AD field begins to oscillate by $V_{\rm gauge}$.
In this case,
of course,
$V_{\rm gauge}\gtrsim V_{\rm grav}$ should be satisfied,
which leads to $\phi_{\rm osc}\lesssim\phi_{\rm eq}$.
Moreover,
$dE/dQ>1{\rm\,GeV}$ leads to 
\begin{equation}
	\frac{\phi_{\rm osc}}{M_{\rm pl}}\lesssim 1.2\times10^{-5}\left(\frac{M_F}{10^6{\rm\,GeV}}\right)^2.
	\label{eq:E/Q>1} 
\end{equation}
From Eq.~(\ref{eq:Q-ball_decay}),
the amplitude of the AD field at its oscillation is constrained as
\begin{equation}
	\frac{\phi_{\rm osc}}{M_{\rm pl}}\lesssim 1.7\times10^{-6}\left(\frac{M_F}{10^6{\rm\,GeV}}\right)^{6/5}.
	\label{eq:Q-ball_decay_condition} 
\end{equation}
The resulting baryon asymmetry is then estimated as
\begin{eqnarray}
	Y_B&=&\frac{n_B}{\rho_{\eta,{\rm BB}}}\frac{\rho_{\eta,{\rm BB}}}{s} \nonumber \\
	&\simeq&1.0
	\left(\frac{\Omega_{\eta,{\rm BB}}h^2}{0.12}\right)
	\left(\frac{\eta_0}{M_{\rm pl}}\right)^{-2}
	\left(\frac{m_{3/2}}{200{\rm\,keV}}\right)
	\left(\frac{M_F}{10^6{\rm\,GeV}}\right)^{-4}
	\left(\frac{\phi_{\rm osc}}{M_{\rm pl}}\right)^4, 
\end{eqnarray}
where
we use Eq.~(\ref{eq:n_B/rho_eta}) and
$H_{\rm osc}\simeq M_F^2/\phi_{\rm osc}$.
Since $\phi_{\rm osc}$ is restricted from above by Eq.~(\ref{eq:Q-ball_decay_condition}),
the baryon asymmetry has an upper bound of
\begin{equation}
	Y_B\lesssim8.5\times10^{-24}
	\left(\frac{\Omega_{\eta,{\rm BB}}h^2}{0.12}\right)
	\left(\frac{\eta_0}{M_{\rm pl}}\right)^{-2}
	\left(\frac{m_{3/2}}{200{\rm\,keV}}\right)
	\left(\frac{M_F}{10^6{\rm\,GeV}}\right)^{4/5}. 
	\label{eq:unstable_gauge}
\end{equation}
Note that Eq.~(\ref{eq:E/Q>1}) and $\phi_{\rm osc}\lesssim\phi_{\rm eq}$ 
are satisfied when $M_F\gtrsim9.0\times10^{4}{\rm\,GeV}$
and $m_{3/2}\lesssim{\cal O}(200){\rm\,keV}$.
One can find that the estimated baryon number is too small to explain the present baryon asymmetry.
This results from the smallness of the amplitude of the AD field
to make Q-balls unstable.

In the above discussion,
we neglect the thermal logarithmic potential.
Next,
we assume that the reheating completes before the moduli oscillation
and that the AD field begins to oscillate by the thermal logarithmic potential
before the reheating.\footnote{
$V^{(2)}_{\rm th}\gtrsim V_{\rm gauge}$ and $V^{(2)}_{\rm th}\gtrsim V_{\rm grav}$
lead to $\phi_{\rm osc}/M_{\rm pl}\lesssim(T_{\rm RH}/M_F)^2$ and 
$\phi_{\rm osc}/M_{\rm pl}\lesssim(T_{\rm RH}^2/m_{3/2}M_{\rm pl})^{1/2}$.
These are satisfied if the reheating completes before the moduli oscillation.
}
Even in this case,
Eq.~(\ref{eq:n_B/rho_eta}) is applicable when $\phi_{\rm osc}$ is near the Planck scale
since the AD field immediately dominates the total energy density after the reheating.
The charge of the Q-ball is given by
$Q\simeq\beta(\phi_{\rm osc}/T_{\rm osc})^4$,
where $T_{\rm osc}$ is the temperature at onset of the AD field oscillation
and is estimated as
$T_{\rm osc}\simeq(M_{\rm pl}T_{\rm RH}^2H_{\rm osc})^{1/4}$,
with $H_{\rm osc}\simeq T_{\rm osc}^2/\phi_{\rm osc}$.
$dE/dQ>1{\rm\,GeV}$ leads to 
\begin{equation}
	\frac{\phi_{\rm osc}}{M_{\rm pl}}\lesssim3.7\left(\frac{M_F}{10^6{\rm\,GeV}}\right)^{2/3}
	\left(\frac{T_{\rm RH}}{6\times10^{11}{\rm\,GeV}}\right)^{2/3}.
	\label{eq:E/Q>1_2} 
\end{equation}
From Eq.~(\ref{eq:Q-ball_decay}),
the amplitude of the AD field at its oscillation is constrained as
\begin{equation}
	\frac{\phi_{\rm osc}}{M_{\rm pl}}\lesssim
	1.0\left(\frac{M_F}{10^6{\rm\,GeV}}\right)^{2/15}
	\left(\frac{T_{\rm RH}}{6\times10^{11}{\rm\,GeV}}\right)^{2/3}.
	\label{eq:Q-ball_decay_condition_2} 
\end{equation}
The resulting baryon asymmetry is estimated as
\begin{eqnarray}
	Y_B&=&\frac{n_B}{\rho_{\eta,{\rm BB}}}\frac{\rho_{\eta,{\rm BB}}}{s} \nonumber \\
	&\simeq&2.9\times10^{-50}
	\left(\frac{\Omega_{\eta,{\rm BB}}h^2}{0.12}\right)
	\left(\frac{\eta_0}{M_{\rm pl}}\right)^{-2}
	\left(\frac{m_{3/2}}{200{\rm\,keV}}\right)
	\left(\frac{T_{\rm RH}}{M_{\rm pl}}\right)^{-4}
	\left(\frac{\phi_{\rm osc}}{M_{\rm pl}}\right)^6. 
\end{eqnarray}
This has an upper bound of\footnote{
By taking into account of the ellipticity parameter $\epsilon\sim m_{3/2}/H_{\rm osc}$,
we obtain more severe upper bounds of Eq.~(\ref{eq:unstable_gauge}) and Eq.~(\ref{eq:unstable_thermal}) 
than the estimated value in~\cite{Kasuya:2001tp}.
}
\begin{equation}
	Y_B\lesssim8.5\times10^{-24}
	\left(\frac{\Omega_{\eta,{\rm BB}}h^2}{0.12}\right)
	\left(\frac{\eta_0}{M_{\rm pl}}\right)^{-2}
	\left(\frac{m_{3/2}}{200{\rm\,keV}}\right) 
	\left(\frac{M_F}{10^6{\rm\,GeV}}\right)^{4/5}, 
	\label{eq:unstable_thermal}
\end{equation}
where we use Eq.~(\ref{eq:Q-ball_decay_condition_2}).
Note that Eq.~(\ref{eq:E/Q>1_2}) is satisfied when $M_F\gtrsim9.0\times10^{4}{\rm\,GeV}$.
The resulting baryon asymmetry is too small to explain the observed baryon asymmetry.
This results from the early oscillation of the AD field by the thermal logarithmic potential.

\section{$LH_u$ flat direction}
\label{sec:LH_u}

In the previous section,
we focus on flat directions which have only soft SUSY breaking terms
and non-renormalizable terms.
However,
there exists an exception.
The $LH_u$ flat direction has the supersymmetric $\mu$-term of Higgs supermultiplets 
in addition to the SUSY breaking terms.
Moreover,
one-loop correction gives a positive contribution ($K>0$).
This implies that
the $LH_u$ flat direction does not form into the Q-ball.

First,
let us consider the case that the $\mu$-term exists before the thermal inflation
and show that it is still difficult to explain the baryon asymmetry.
Assuming that the AD field begins to oscillate because of the $\mu$-term
before the thermal inflation,
the ratio of lepton number to Big-Bang moduli density is given by
\begin{equation}
	\left|\frac{n_L}{\rho_{\eta,{\rm BB}}}\right|=\frac{2m_{3/2}}{\mu^2}
	\left(\frac{\phi_{\rm osc}}{M_{\rm pl}}\right)^2\left(\frac{\eta_0}{M_{\rm pl}}\right)^{-2},
	\label{eq:n_L/rho_eta}
\end{equation}
where $\mu$ is the electroweak scale.
Here,
we use Eq.~(\ref{eq:n_B/rho_eta})
which is applicable when the total energy of the universe scales as $a^{-3}$
between the eras when the AD field starts to oscillate and when the moduli field starts to oscillate.
Even if the reheating completes before the onset of the moduli oscillation,
the AD field immediately dominates the total energy of the universe
when $\phi_{\rm osc}\simeq M_{\rm pl}$
and the above estimation is valid in that case.
The lepton asymmetry is partially converted into baryon asymmetry through the sphaleron process.
The relation between the lepton and baryon number is given by
\begin{equation}
	\frac{n_B}{s}=-\frac{8}{23}\frac{n_L}{s}.
\end{equation}
The produced baryon asymmetry is estimated as
\begin{equation}
	Y_B\simeq 6.1\times10^{-20}\left(\frac{\Omega_{\eta,{\rm BB}}h^2}{0.12}\right)
	\left(\frac{\eta_0}{M_{\rm pl}}\right)^{-2}
	\left(\frac{m_{3/2}}{200{\rm\,keV}}\right)
	\left(\frac{\mu}{1{\rm\,TeV}}\right)^{-2}
	\left(\frac{\phi_{\rm osc}}{M_{\rm pl}}\right)^2, 
\end{equation}
and the resulting asymmetry is too small to explain the present asymmetry.
This results from the early oscillation of the AD field by the $\mu$-term
whose scale is much larger than the gravitino mass scale,
$\mu\gg m_{3/2}$.

Hereafter,
we consider an alternative scenario where
the $\mu$-term is negligible at the onset of the oscillation of the AD field
and is generated at the end of the thermal inflation.
The coupling between the flaton and Higgs supermultiplets (see Eq.~(\ref{eq:mu-term}))
prohibits the $\mu$-term before the thermal inflation
and plays the role of generating the $\mu$-term by the VEV of the flaton after the thermal inflation.
In this case,
the scenario changes as follows.
The $LH_u$ flat direction produces lepton asymmetry by soft SUSY breaking terms
and forms into Q-balls before the thermal inflation.
After the thermal inflation,
the generated $\mu$-term violates the condition for the existence of the Q-ball.
Q-balls decay and the absorbed lepton number is released.
In the following,
we show if the released lepton number can explain the observed baryon asymmetry.

In order to generate the $\mu$-term,
we assume that the flaton $X$ couples with the Higgs supermultiplets
as Eq.~(\ref{eq:mu-term}).
Then,
the following $F$-term potential arises:
\begin{equation}
	V_{F}=\frac{\left|\lambda_{\mu}\right|^2|X|^4}{M_{\rm pl}^2}|\Phi|^2\equiv\mu^2(X)|\Phi|^2,
	\label{eq:mu-term_potential}
\end{equation}
where $\Phi$ is the up-type Higgs scalar field, namely the AD field.
For convenience,
we introduce the $\mu$ parameter defined as Eq.~(\ref{eq:mu-term_potential}).
Then,
the flaton potential is expressed as
\begin{equation}
	V(X)\simeq V_0+\left(\frac{\alpha^*\lambda_X}{M_{\rm pl}}X^3+{\rm h.c.}\right)+\frac{|\lambda_{\mu}|^2|\Phi|^2}{M_{\rm pl}^2}|X|^4+\frac{|\lambda_X|^2}{M_{\rm pl}^2}|X|^6,
	\label{eq:X_potential}
\end{equation}
where we neglect the linear terms and the quartic terms in Eq.~(\ref{eq:modified_potential})
(see the discussion in Sec.~\ref{subsec:linear_term}).
Before the thermal inflation,
the VEV of $X$ is so small that
$\mu$-term is negligible for the dynamics of the AD field
compared with SUSY breaking terms,
which implies that the AD field forms into Q-balls.
Then,
the flaton acquires the VEV after the thermal inflation,
which gives the $\mu$-term to the AD field potential.
If it breaks the condition for the existence of the Q-ball,
the Q-ball decays.
The lepton charge is then released to the thermal plasma.

The $\mu$ parameter must increase from the outside to the inside of the Q-ball
in order to violate the condition for the existence of the Q-ball.
Namely,
the following relation must be realized:
\begin{equation}
	\frac{\mu^2(X(|\Phi_{\rm in}|))}{\mu^2(X(|\Phi_{\rm out}|))}>1.
	\label{eq:mu_condition}
\end{equation}
$\Phi_{\rm in}$ and $\Phi_{\rm out}$ show the AD field values inside and outside the Q-ball,
respectively.
From Eq.~(\ref{eq:mu-term_potential}),
one can find that
the $\mu$ parameter explicitly depends on $|X|$.
$|X|$ is determined by the potential of Eq.~(\ref{eq:X_potential}),
which depends on $|\Phi|$ via the interaction of Eq.~(\ref{eq:mu-term_potential}).
Thus,
one can find that
the $\mu$ parameter depends on the AD field value.

Since $|\Phi|\neq0$ inside the Q-ball,
the coupling term between the flaton and the AD field lifts the flaton potential
and $|X|$ inside the Q-ball is smaller than that outside the Q-ball.
Hence,
Eq.~(\ref{eq:mu_condition}) is not satisfied at the tree level.
Taking one-loop corrections into account,
however,
we find that the $\mu$-term can be steeper than a quadratic mass term for a larger VEV of the AD field
$|\Phi|$.
It is expressed as
\begin{equation}
	V=\left[\frac{\left|\lambda_{\mu}\right|^2M^4}{M_{\rm pl}^2}\left(1+K\log\frac{|\Phi|^2}{M_*^2}\right)\right]|\Phi|^2,
\end{equation}
where $K>0$ and $|K|\simeq{\cal O}(0.1\mathchar`-0.01)$.
The parameter $M$ is the VEV of the flaton.
For different amplitudes of the AD field ($|\Phi_+|>|\Phi_-|$),
the ratio of the $\mu$-term is estimated as
\begin{equation}
	\frac{\mu^2(|\Phi_{+}|)}{\mu^2(|\Phi_{-}|)}
	\simeq\frac{M_+^4}{M_-^4}
	\left(1+K\log\frac{|\Phi_{+}|^2}{|\Phi_{-}|^2}\right),
\end{equation}
where $M_+$ and $M_-$ are the flaton field values at $|\Phi_+|$ and $|\Phi_-|$,
and satisfy $M_+<M_-$.
Precisely speaking,
when the second term in the parenthesis is ${\cal O}(1)$,
the perturbation breaks down
and one should solve renormalization group equations.
For simplicity,
we assume that the one-loop correction factor is 2 at most\footnote{
Even if we assume that the correction factor is larger than 2,
our result does not change significantly.
}
and require 
\begin{equation}
	\frac{1}{2}<\frac{M_{\rm in}^4}{M_{\rm out}^4}<1
	\label{eq:M_in/M_out}
\end{equation}
to realize the condition of Eq.~(\ref{eq:mu_condition}).
Here,
$M_{\rm in}$ and $M_{\rm out}$ correspond to the flaton VEVs
inside and outside the Q-ball.

In order for the flaton VEV not to be highly damped inside the Q-ball,
the quartic term of the flaton in Eq.~(\ref{eq:X_potential}) should be small enough.
Thus,
the AD field value in the Q-ball, $\phi_0$,
which is proportional to the oscillation amplitude in the case of the gauge-mediation type Q-ball,
should be small.
On the other hand,
in order to produce huge baryon number,
$\phi_{\rm osc}$ should be large.
Hence,
we focus on the delayed-type Q-ball
because $\phi_0$ is independent of $\phi_{\rm osc}$.
$\phi_0$ is determined by the field value,
where the gauge mediation effect is comparable to the gravity mediation effect,
and is estimated as
\begin{equation}
	\phi_0\simeq \frac{1}{\sqrt{2}}\beta^{1/4}\phi_{\rm eq}\simeq
	3.6\times10^{13}{\rm\,GeV}\left(\frac{m_{3/2}}{200{\rm\,keV}}\right)^{-1}
	\left(\frac{M_F}{4\times10^{4}{\rm\,GeV}}\right)^2.
\end{equation}
Substituting this estimated value into the flaton potential of Eq.~(\ref{eq:X_potential}),
the flaton VEV inside the Q-ball is obtained by solving the following equation:
\begin{equation}
	-3\frac{|\alpha||\lambda_X|}{M_{\rm pl}}M_{\rm in}^2
	+2\frac{|\lambda_{\mu}|^2\phi_0^2}{M_{\rm pl}^2}M_{\rm in}^3
	+3\frac{|\lambda_X|^2}{M_{\rm pl}^2}M_{\rm in}^5=0.
	\label{eq:M_in}
\end{equation}
In order to estimate the parameter $\alpha$,
we introduce a dimensionless parameter $\zeta$ as follows:
\begin{equation}
	|\alpha|\equiv\zeta\frac{|\lambda_X|M_{\rm in}^3}{M_{\rm pl}}.
	\label{eq:zeta}
\end{equation}
Note that $\zeta=1$ corresponds to the case without the coupling of Eq.~(\ref{eq:mu-term_potential})
and that $\zeta>1$ is satisfied.
Using Eqs.~(\ref{eq:M_alpha}) and (\ref{eq:zeta}),
one can find that the ratio of the flaton VEV inside to outside the Q-ball is given by
\begin{equation}
	\frac{M_{\rm in}}{M_{\rm out}}=\frac{1}{\zeta^{1/3}},
\end{equation}
and the condition of Eq.~(\ref{eq:M_in/M_out}) leads to $1<\zeta<1.7$.
By solving the equation of Eq.~(\ref{eq:M_in}),
one can obtain
\begin{equation}
	M_{\rm in}^2=\frac{2|\lambda_{\mu}|^2\phi_0^2}{3(\zeta-1)|\lambda_X|^2}
\end{equation}
in terms of $\zeta$.
Then,
from Eq.~(\ref{eq:zeta}),
the parameter $\alpha$ can be estimated as
\begin{equation}
	|\alpha|=\zeta\left[\frac{2}{3(\zeta-1)}\right]^{3/2}
	\frac{|\lambda_{\mu}|^3\phi_0^3}{|\lambda_X|^2M_{\rm pl}}.
\end{equation}
By using the expression of the $\mu$-term as Eq.~(\ref{eq:mu-term_potential})
and $1<\zeta<1.7$,
one can obtain the following constraint on the parameter $\alpha$:
\begin{eqnarray}
	|\alpha|&=&\zeta^{1/3}\left[\frac{2}{3(\zeta-1)}\right]^{1/2}\mu(|\Phi_{\rm out}|)\phi_0 \nonumber \\
	&>&1.3\times10^{16}{\rm\,GeV}^2\left(\frac{\mu(|\Phi_{\rm out}|)}{300{\rm\,GeV}}\right)
	\left(\frac{\phi_0}{3.6\times10^{13}{\rm\,GeV}}\right).
\end{eqnarray}
Hence,
when the symmetry breaking parameter $\alpha$ satisfies the above constraint,
the condition of Eq.~(\ref{eq:mu_condition}) can be realized including one-loop corrections.
Hereafter,
we assume the $Z_4$ symmetry breaking term of the order of
$|\alpha|\simeq1.3\times10^{16}{\rm\,GeV}^2$.

From Eq.~(\ref{eq:M_alpha}),
the energy of the thermal inflation is estimated as\footnote{
Since $(V_0/3M_{\rm pl}^2)^{1/2}\gtrsim m_{\eta}$,
the moduli field may start to oscillate
after the potential energy of the flaton dominate the energy of the universe,
which implies that the thermal inflation cannot work.
Here,
we assume that the moduli field starts to oscillate
before the thermal inflation begins.
}
\begin{equation}
	V_0\simeq 1.7\times 10^{32}{\rm\,GeV}^4\left(\frac{|\alpha|}{1.3\times10^{16}{\rm\,GeV}^2}\right)^2.
\end{equation}
The flaton and $R$-axion mass is estimated as
\begin{equation}
	m_{\chi}\simeq m_{a}\simeq 450{\rm\,GeV}\left(\frac{|\alpha|}{1.3\times10^{16}{\rm\,GeV}^2}\right)^{2/3}\left(\frac{|\lambda_X|}{5\times10^{-8}}\right)^{1/3},
\end{equation}
where the coupling constant $|\lambda_X|$ is assumed to be small enough 
for the flaton mass to be smaller than sparticle mass,
which keeps it from decaying into sparticle pairs.
From Eq.~(\ref{eq:M_alpha}),
the VEV of the flaton at the true minimum is given by
\begin{equation}
	M\simeq 8.6\times10^{13}{\rm\,GeV}\left(\frac{|\alpha|}{1.3\times10^{16}{\rm\,GeV}^2}\right)^{1/3}
	\left(\frac{|\lambda_X|}{5\times10^{-8}}\right)^{-1/3}.
\end{equation}
Then,
the $\mu$-term outside the Q-ball is given by
\begin{equation}
	\mu(|\Phi_{\rm out}|)\simeq 300{\rm\,GeV}\left(\frac{|\lambda_{\mu}|}{10^{-7}}\right)
	\left(\frac{M}{8.6\times10^{13}{\rm\,GeV}}\right)^2,
\end{equation}
where the coupling constant $|\lambda_{\mu}|$ is also assumed to be as small as $10^{-7}$
to obtain the $\mu$-term of the electroweak scale.

After the thermal inflation,
the VEV of the flaton violates the condition for the existence of the Q-ball
and Q-balls decay. 
The released lepton number is converted into baryon number
through the sphaleron process.
The provided baryon asymmetry is estimated as
\begin{eqnarray}
	Y_B&=&\frac{8}{23}
	\left.\frac{-n_L}{\rho_{\eta,{\rm BB}}}\right|_{\eta~{\rm osc}}
	\left.\frac{\rho_{\eta,{\rm BB}}}{s}\right|_{0} \nonumber \\
	&\simeq&7.7\times10^{-10}
	\left(\frac{T_{\rm RH}}{5\times10^7{\rm\,GeV}}\right)\left(\frac{m_{3/2}}{200{\rm\,keV}}\right)^{-1} \nonumber \\
	&&\times\left(\frac{V_0}{1.7\times10^{32}{\rm\,GeV}^4}\right)^{-1}
	\left(\frac{T^{\chi}_{\rm RH}}{10{\rm\,MeV}}\right)
	\left(\frac{T_c}{20{\rm\,TeV}}\right)^3\left(\frac{\phi_{\rm osc}}{M_{\rm pl}}\right)^2.
\end{eqnarray}
One can find that the observed baryon asymmetry,
$Y_{B}\simeq 8\times10^{-11}$,
could be explained if the AD field begins to oscillate near the Planck scale.
Q-balls collapse irrespective of the oscillation amplitude of the AD field,
which is contrary to the case of unstable Q-balls in Sec.~\ref{subsec:unstable}.

Next,
we estimate the temperature at the end of the thermal inflation, $T_c$,
and the reheating temperature of the flaton decay, $T^{\chi}_{\rm RH}$.
The thermal inflation ends when $T_c\sim c^{1/2}\left\langle|X|\right\rangle$
because thermal particles which couple with the flaton become massive.
From Eq.~(\ref{eq:local_minimum}),
the temperature at the end of the thermal inflation, $T_c$, is estimated as
\begin{equation}
	T_c\simeq 17{\rm\,TeV}\left(\frac{|\alpha|}{1.3\times10^{16}{\rm\,GeV}^2}\right)^{1/3}
	\left(\frac{m_{3/2}}{200{\rm\,keV}}\right)^{1/3},
\end{equation}
where we assume $c\simeq{\cal O}(1)$.
Note that the released lepton number is successfully converted to baryon number through the sphaleron process
since the thermal inflation ends before the electroweak symmetry breaking.
As for the reheating temperature,
$T^{\chi}_{\rm RH}$ is estimated from the decay rate of the flaton.
Since $m_{\chi}>2m_h$,
the flaton mainly decays into two Higgs bosons at the tree level
and they decay into the Standard Model particles.
From Eq.~(\ref{eq:Gamma_2h}),
The reheating temperature,
$T^{\chi}_{\rm RH}$,
is expressed as
\begin{eqnarray}
	T^{\chi}_{\rm RH}&\simeq&\left(\frac{90}{\pi^2 g_*}\right)^{1/4}\sqrt{\Gamma(\chi\to2h)M_{\rm pl}} 
	\nonumber \\
	&\simeq&11{\rm\,MeV}
	\left(\frac{\mu(|\Phi_{\rm out}|)}{300{\rm\,GeV}}\right)^2
	\left(\frac{m_{\chi}}{450{\rm\,GeV}}\right)^{-1/2}
	\left(\frac{M}{8.6\times10^{13}{\rm\,GeV}}\right)^{-1},
\end{eqnarray}
where we use $g_*=10.8$.
Hence,
the reheating temperature can be higher than ${\cal O}(10){\rm\,MeV}$,
which can avoid spoiling the success of the BBN.

We turn to estimate the density parameter of the moduli.
The density parameter of the Big-Bang moduli after the entropy dilution is given by
\begin{eqnarray}
	\Omega_{\eta,{\rm BB}}h^2&\simeq&6.1\times10^{-5}\left(\frac{T_{\rm RH}}{5\times10^7{\rm\,GeV}}\right)
	\left(\frac{V_0}{1.7\times10^{32}{\rm\,GeV}^4}\right)^{-1} \nonumber \\
	&&\times\left(\frac{T_{\rm RH}^{\chi}}{10{\rm\,MeV}}\right)\left(\frac{T_c}{20{\rm\,TeV}}\right)^3
	\left(\frac{\eta_0}{M_{\rm pl}}\right)^2.
\end{eqnarray}
It is found that the Big-Bang moduli is sufficiently diluted.
On the other hand,
the density parameter of the thermal inflation moduli is given by
\begin{equation}
	\Omega_{\eta,{\rm TH}}\simeq 8.1\times10^{7} c_H^2\left(\frac{V_0}{1.7\times10^{32}{\rm\,GeV}^4}\right)\left(\frac{m_{3/2}}{200{\rm\,keV}}\right)^{-2}\left(\frac{T^{\chi}_{\rm RH}}{10{\rm\,MeV}}\right)
	\left(\frac{\eta_0}{M_{\rm pl}}\right)^{2}.
\end{equation}
One can find that the density of the thermal inflation moduli 
is much larger than the critical density.
Therefore,
in order for this scenario to work,
the separation between the local minimum determined by the Hubble induced term
and the true minimum should be of the order of $10^{-4}M_{\rm pl}$.\footnote{
If the moduli field has no couplings with photons,
the constraint from the X-ray background spectra is irrelevant.
In this case,
the moduli field with mass of ${\cal O}(1){\rm\,MeV}$,
which cannot decay into electrons,
could be the dark matter
and the fine-tuning can be relaxed.
However,
it is still difficult for this scenario to work unless fine-tuning is allowed.
}
Although this may result from $0.01\%$ fine-tuning of the moduli potential,
we have no motivation for the thermal inflation in that case
since the moduli problem can be solved 
when $T_{\rm RH}\simeq 10{\rm MeV}$ and $\eta_0/M_{\rm pl}\simeq{\cal O}(10^{-3})$
(see Eq.~(\ref{eq:BB_entropy_1})).
Therefore,
it is found that this scenario cannot work due to the secondary produced moduli.

\section{Summary and Discussions}
\label{sec:}

In gauge-mediated SUSY breaking models,
the cosmic density of the coherently oscillating moduli
exceeds the dark matter density 
unless it is diluted by some entropy production mechanism after the primordial inflation.
The thermal inflation can successfully dilute 
the moduli abundance.
However,
there exists another fatal problem.
The thermal inflation also dilutes baryon number
which may be produced beforehand.
Hence,
some viable baryogenesis is needed
for a cosmologically consistent scenario.
A promising candidate for a mechanism to produce huge baryon asymmetry is the Affleck-Dine mechanism.
In this paper,
we have studied if the thermal inflation is consistent with
the Affleck-Dine baryogenesis in gauge-mediated SUSY breaking models.

Q-ball formation is inevitable for the logarithmic potential
induced from the gauge-mediated SUSY breaking effects
and becomes a hinderance to the baryon number production.
In order to produce huge baryon number,
the amplitude of the AD field should be large,
which renders the formed Q-balls stable.
It is difficult to extract baryon charge from stable Q-balls and
the evaporated baryon charge cannot explain the observed baryon asymmetry.
On the other hand,
unstable Q-balls can release all baryon charge.
In this case,
however,
the amplitude of the AD field is restricted from above
in order to prohibit Q-ball decay during and after the BBN epoch.
Thus,
sufficient baryon number cannot be produced.
As a result,
the Affleck-Dine mechanism is incompatible with the thermal inflation to solve
the moduli problem in both cases.

Among the flat directions,
the $LH_u$ flat direction is special
since it has the supersymmetric $\mu$-term
which violates the condition for the existence of the Q-ball.
Since the flaton can have a coupling with the the Higgs supermultiplets,
the flaton VEV naturally generates the $\mu$-term.
We showed that the generated $\mu$-term at the end of the thermal inflation
destroys formed Q-balls 
and that the released lepton asymmetry could explain the baryon asymmetry.
In this case,
however,
the energy of the thermal inflation $V_0$ is required to be large
in order to violate the condition for the existence of the Q-ball,
which leads to overproduction of the thermal inflation moduli.
Thus,
the moduli problem arises again because of the secondary produced moduli relic
and serious fine-tuning is needed for this scenario to work.
We conclude that
it is difficult for the thermal inflation to be consistent with the Affleck-Dine mechanism.

One might think that the electroweak baryogenesis can explain the asymmetry without dilution.
In that case,
the reheating temperature after the thermal inflation,
$T^{\chi}_{\rm RH}$,
should be higher than the electroweak scale.
However,
it seems to be difficult to dilute the moduli density 
for such high reheating temperature in the context of the thermal inflation.

Finally,
we make some comments on gravity-mediated SUSY breaking models.
In these models,
the gravitino mass is comparable to sparticle mass of the order of $m_{\rm SUSY}\simeq{\cal O}(1){\rm\,TeV}$.
The usual Affleck-Dine mechanism cannot provide sufficient baryon number to survive after the dilution
because of the early oscillation of the AD field by the soft SUSY breaking mass term.
Instead,
the modified Affleck-Dine mechanism,
which has been proposed by~\cite{Stewart:1996ai},
could explain the observed baryon asymmetry.
In this model,
the $LH_u$ flat direction provides lepton number after the thermal inflation.
The dynamics of the $LH_u$ flat direction is so complicated 
that numerical simulations are necessary.
Some works have revealed that this modified mechanism can work 
in the gravity-mediated SUSY breaking models~\cite{Jeong:2004hy,Kawasaki:2006py},
but it is unclear if it can also work well in gauge-mediated SUSY breaking models.

\section*{Acknowledgments}
This work is supported by Grant-in-Aid for Scientific research 
from the Ministry of Education, Science, Sports, and Culture
(MEXT), Japan,
No.~15H05889 (M.K.) and No.~25400248 (M.K.), 
JSPS Research Fellowships for Young Scientists (No.~25.8715 (M.Y.)), 
World Premier International Research Center Initiative (WPI Initiative), MEXT, Japan (M.K. and M.Y.),
and the Program for the Leading Graduate Schools, MEXT, Japan (T.H. and M.Y.).


\end{document}